\documentclass[10pt]{elsart}
\usepackage{amssymb}
\usepackage[dvips]{graphicx}

\begin{document}
\begin{frontmatter}
\title{Search for $\mathbf{d^*}$ Dibaryon by Double-radiative Capture on Pionic Deuterium}
\author[UKY]{P.~A.~{\.Z}o{\l}nierczuk}, 
\author[UKY]{T.~P.~Gorringe}, 
\author[UBC]{M.~D.~Hasinoff},
\author[UKY]{M.~A.~Kovash},
\author[UKY]{S.~Tripathi}, 
\author[TRI]{D.~H.~Wright\thanksref{SLAC}}
\address[UKY]{University of Kentucky, Lexington, KY 40506, USA}
\address[UBC]{University of British Columbia, Vancouver, B.C., Canada V6T 1Z1}
\address[TRI]{TRIUMF, 4004 Wesbrook Mall, Vancouver, B.C., Canada V6T 2A3} 
\thanks[SLAC]{Present address: SLAC, P.O. Box 20450, Stanford, CA 94309, USA}

\begin{abstract}
We report a search 
for $d^*$ dibaryon production by double-radiative capture
on pionic deuterium. 
The experiment was conducted 
at the TRIUMF cyclotron
using the RMC cylindrical pair spectrometer,
and detected $\gamma$-ray coincidences
following  pion stops in liquid deuterium.
We found no evidence for narrow dibaryons,
and obtained a branching ratio upper limit, $BR < 6.7 \times 10^{-6}$ (90\% C.L.),
for narrow $d^*$ production 
in the mass range from $1920$ to $1980$~MeV.
\end{abstract}

\begin{keyword} {dibaryon} \sep {double-radiative capture} \sep{pionic deuterium} 
\PACS{14.20.Pt, 25.80.Hp, 36.10.Gv}
\end{keyword}
\end{frontmatter}

\section{Introduction}
At present the deuteron 
is the only established particle
with a baryon number $B = 2$.
However a large number of theoretical predictions for additional dibaryons 
have been published in the literature. 
The predictions have 
exploited both quark--gluon and hadron viewpoints,
and include objects such as exotic multi--quark states 
and molecular--like baryonic states.
Obviously the experimental discovery of another dibaryon
would provide new insight into hadron 
and quark--gluon dynamics at the GeV scale.  

A recent claim for dibaryon production 
has been published 
by the Di2$\gamma$ collaboration 
at the JINR phasotron \cite{Kh01}. 
The collaboration measured 
the proton--proton double bremsstrahlung reaction
$p p \rightarrow p p \gamma \gamma$
by directing 216~MeV protons onto a liquid hydrogen target 
and recording $\gamma$--ray coincidences in two CsI/NaI detector arrays.
The authors observed an intriguing structure
in the background--subtracted energy spectrum
of the $\gamma$--ray coincidence data.
It comprised a relatively narrow peak centered at about 24~MeV
and a relatively broad peak centered at about 60~MeV.
They attributed the structure 
to the $d^*$ dibaryon with mass $1956 \pm 6$~MeV and width $\leqslant 8$~MeV.
They hypothesized that $d^*$ dibaryons
were first produced via the two-body process $p p \rightarrow d^* \gamma$
and then decayed via the three-body process $d^* \rightarrow p p \gamma$,
thus contributing to $p p$ double bremsstrahlung.

The $d^*$ dibaryon of Khrykin {\it et al.}\ \cite{Kh01}
would have an electric charge $Q = +2$,
an isospin component $T_z = +1$,
and therefore an isospin of $T \geqslant 1$.
The narrow width of the claimed $d^*$ state implies that
its quantum numbers either forbid or strongly suppress the $NN$ decay mode.
Thus, if the $d^*$ is isovector ($T=1$),
it must have spin-parities $J^{\pi} = 1^+, 3^+, 5^+$ etc.
Moreover Khrykin {\it et al.}\ have argued 
that $( J^{\pi} , T ) = ( 1^+ , 1 )$
is the most natural choice
for the $d^*$'s quantum numbers, 
being the lowest spin--isospin pp-decoupled state
with zero orbital angular momentum.

Recently Gerasimov \cite{Ge98,Ge01} suggested 
that double-radiative capture on pionic deuterium
\begin{equation}
\label{e:pid}
\pi^- d \rightarrow n n \gamma \gamma
\end{equation}
is an excellent candidate for further investigations
of the dibaryon's existence.
In Eqn.~\ref{e:pid} the $d^* \; (T_z=-1)$ dibaryon 
is first produced via radiative capture $\pi^- d \rightarrow \ d^* \gamma$
and then disintegrates via radiative decay $d^* \rightarrow n n \gamma$.
Using a simple model 
Gerasimov estimated\footnote{Gerasimov 
assumed a  N$\Delta$ bound state for the $d^*$ dibaryon
and a Kroll-Ruderman graph $\pi N \rightarrow \gamma \Delta$  
for the production mechanism.}
that the branching ratio
for the $d^*$--mediated process
might be as large as 0.5\%.
This yield would exceed 
by $100$ times
the expected two--photon branching ratio
for non--resonant double-radiative capture
in pionic deuterium
(see \cite{Ge98,Zo00}).

A number of $\gamma$--ray experiments on pionic deuterium
have already been conducted, 
so is it possible for the signatures of the $d^*$ dibaryon to have been missed?
Singles $\gamma$--ray data 
on pionic deuterium 
are available
from the TRIUMF experiments of Highland {\it et al.}\ \cite{Hi81}
and Stanislaus {\it et al.}\ \cite{St89} and 
the PSI experiment of Gabioud {\it et al.}\ \cite{Ga84}.
Unfortunately, because 
of the large branching ratio $BR = 0.26$
for the single radiative capture reaction,
which yields
a $\sim$130~MeV $\gamma$--ray peak 
with a large low--energy tail,
a small contribution 
from dibaryon production
with a branching ratio $\leqslant$10$^{-2}$ 
is not excluded 
(the exact sensitivity of the singles $\gamma$--ray experiments
is a strong function of the mass and the width of the dibaryon).
Coincidence $\gamma$--ray data
on pionic deuterium 
are also available from the TRIUMF experiment
of MacDonald {\it et al.}\ \cite{Ma77}. 
However this experiment was designed
for back--to--back photons originating
from $\pi^o \rightarrow \gamma \gamma$ decay 
following $^2$H$( \pi^- , \pi^o )$ charge exchange.
Consequently their sensitivity 
to two-photon events from $d^*$ dibaryon production
in double-radiative capture was very low.

\section{Experimental Setup}
\begin{figure}
\begin{center}
\includegraphics[height=0.6\hsize]{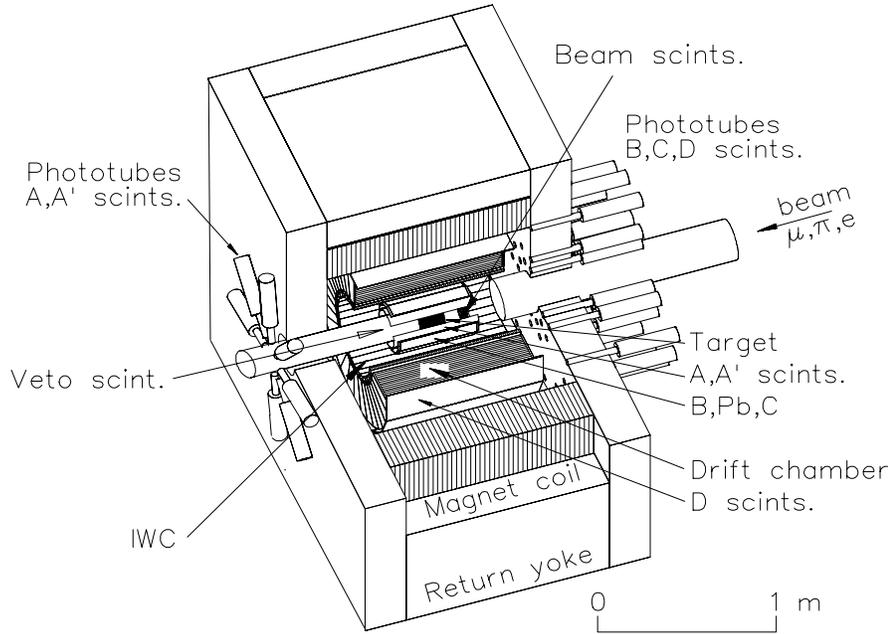}
\end{center}
\caption{The RMC spectrometer showing
the deuterium target, lead converter, 
cylindrical drift chamber, 
trigger scintillators
and spectrometer magnet.} 
\label{fig:detector}
\end{figure}

Herein we report a dedicated search for dibaryon production
via double-radiative capture on pionic deuterium.
The experiment was conducted
on the M9A beamline at the TRIUMF cyclotron 
using the RMC pair spectrometer (see Fig.~\ref{fig:detector}).

The beamline delivered 
a negative pion flux of $5 \times 10^5 \, s^{-1}$
with a central momentum of 81.5~MeV/c
and a momentum bite $\Delta p/p$ of about 10\%.
The beam was unseparated,\footnote{the RF separator was unavailable for the data taking} 
having a $e/\pi$ ratio of about 14/1
and a $\mu/\pi$ ratio of about 1/1.
The incoming pions 
were counted in a 4-element plastic scintillator telescope
and stopped in a 2.5~liter liquid deuterium target.

The outgoing photons were detected 
by electron-positron pair production in a 1~mm cylindrical lead converter 
and $e^+$, $e^-$ tracking in a cylindrical drift chambers.
A 1.2~kG axial magnetic field was used for momentum analysis
and concentric plastic scintillator rings were used for fast triggering.
The trigger scintillator package consisted of
the A-ring (just inside the Pb converter radius),
the C-ring (just inside the multiwire drift chamber radius),
and the D-ring (just outside the drift chamber radius).
For more information on the spectrometer
see Wright {\it et al.}~\cite{Wr92},
note that in this experiment
we moved the Pb converter
from just inside the C-counter radius
to just outside the A-counter radius.

The two--photon trigger 
was based upon 
the multiplicities and topologies of hits 
in the trigger scintillators
and the drift chamber cells.
The scintillator trigger required 
zero hits in the A-ring,
two or more hits in the C-ring,
three or more hits in the D-ring,
and a C-D topology consistent
with the conversion of two $\gamma$-rays.
The drift chamber trigger imposed
minimum values for number of the drift cell hits
or the drift cell clusters 
in each drift chamber layer.

During a four week running period 
we accumulated $\gamma$-ray coincidence spectra
from a total of about $3.8 \times 10^{11}$ pion stops 
in the deuterium target.
We also collected data
from $\pi^-$ stops in liquid H$_2$
for setup and calibration.

\section{Data Reduction}
The significant backgrounds involved 
(i) true coincidences 
from $\pi^0 \rightarrow \gamma \gamma$ decay
following $(\pi^- , \pi^o )$ charge exchange  
on $^1$H contamination,
(ii) accidental coincidences
from two $\pi^- d \rightarrow n n \gamma$ events
following two pion stops   
in one beam pulse (hereafter denoted $\pi$-$\pi$ events),\footnote{The 
pion beam had a micro--structure with a pulse width of 2--4~ns
and a pulse separation of 43~ns.
For an incident flux of $5 \times 10^{5}$~s$^{-1}$
the probability for more than one pion arriving in a single beam pulse
was about 1.1\%.}
and (iii) accidental coincidences
between a delayed radiative $\mu$ decay $\gamma$-ray
and a prompt radiative $\pi$ capture $\gamma$-ray
(hereafter denoted $\mu$-$\pi$ events).
The $\pi^o$ decay background yielded 
photon-pairs with opening angles of $\cos\theta_{12}  \leqslant -0.76$
and summed energies of $E_{\mathrm{sum}} = E_{\gamma 1} + E_{\gamma 2} \simeq m_{\pi}$.
The accidental coincidence backgrounds yielded 
photon-pairs with opening angles of $-1.0 \leqslant \cos\theta_{12}  \leqslant +1.0$
and summed energies 
of up to $\sim 180$~MeV for $\mu$-$\pi$ accidentals
and $\sim 260$~MeV for $\pi$-$\pi$ accidentals.

\begin{figure}
\begin{center} 
\includegraphics[width=0.7\hsize]{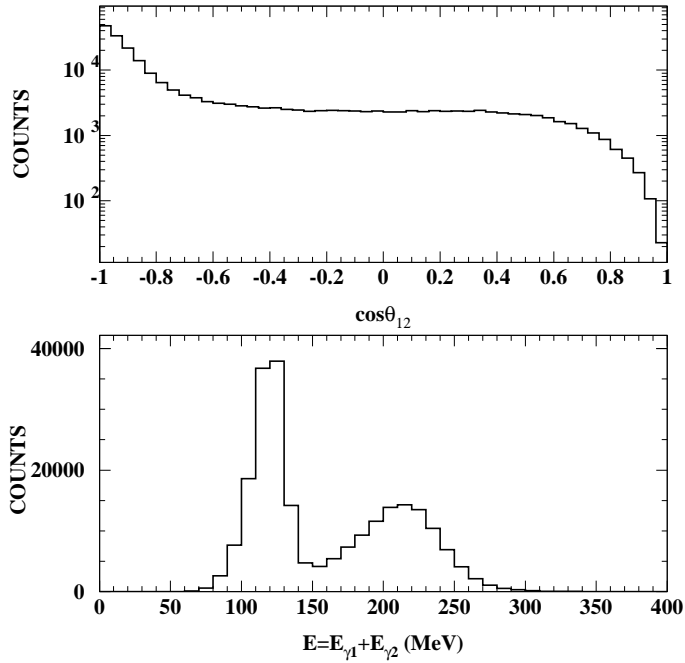}
\end{center}
\caption{The photon opening angle $\cos\theta_{12}$ (top) 
and summed photon energy spectra (bottom)
for events passing the tracking cuts and photon cuts.
The events are mainly
real $\gamma$-$\gamma$ coincidences from $\pi^o$ decay
and accidental $\gamma$-$\gamma$ coincidences from $\pi$-$\pi$
and $\mu$-$\pi$ events.}
\label{fig:raw}
\end{figure}

In analyzing the data a number of cuts were applied 
to select the two-photon events.
A tracking cut 
imposed minimum values for the number of points in the tracks
in the drift chamber and maximum values for the chi--squared of fits to the tracks.
A photon cut 
required that the electron-positron pairs intersect at the lead converter
and that the photon pairs originate from the deuterium target.
The total number of photon pairs,
{\it i.e.} events surviving 
the tracking cuts and photon cuts,
was $2.3\times 10^{5}$.
These photon pairs
(shown in Fig.~\ref{fig:raw})
are dominated 
by the real $\gamma$-$\gamma$ coincidences from $\pi^o$ decays
and accidental $\gamma$-$\gamma$ coincidences from 
$\pi$-$\pi$ events and $\mu$-$\pi$ events.
The accidental $\gamma$-$\gamma$ background
is clearly seen in the summed energy spectrum
as events with $E > 150$~MeV.
The $\pi^o$ decay background
is clearly seen in the opening angle spectrum
as events with $\cos\theta_{12}   < -0.6$. 
In order to remove these backgrounds, 
a beam counter amplitude cut was applied to 
reject the  $\pi$-$\pi$ accidentals
and a C-counter timing cut to 
reject the $\mu$-$\pi$ accidentals.
Finally an opening angle cut 
was used to reject the background 
from $\pi^o$ decay.

\begin{figure}
\begin{center} 
\includegraphics[width=0.7\hsize]{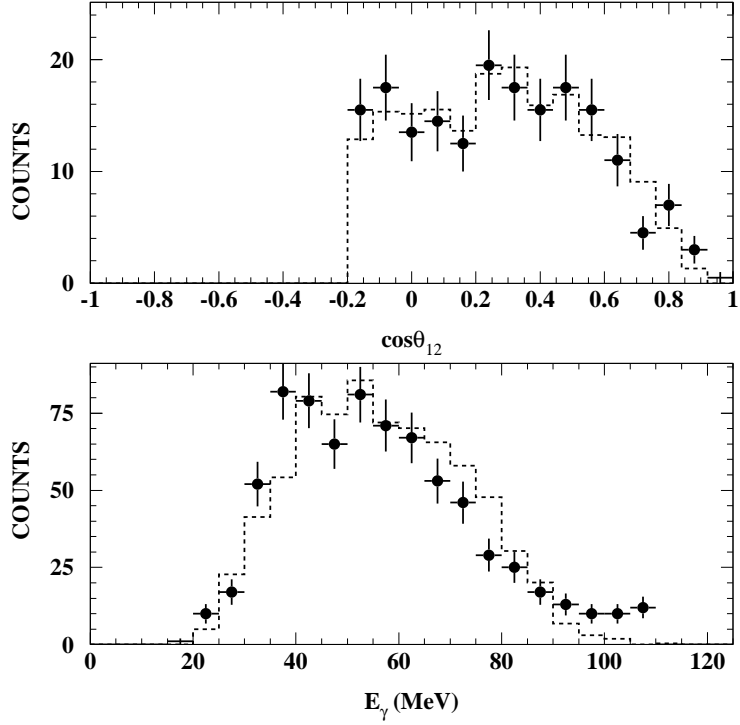}
\end{center}
\caption{The photon opening angle $\cos\theta_{12}$ (top) 
and individual photon energy spectra (bottom) 
for events passing all cuts. Our experimental data are denoted
by full circles and the dashed lines represent Monte Carlo simulations
assuming non-resonant pion double-radiative capture on deuterium.} 
\label{fig:signal}
\end{figure}

A total of 370 two-photon events were found to survive
the beam counter amplitude cut, 
C-counter timing cut, 
and photon opening angle cut.
\footnote{Actually a small contribution of $22 \pm 5$ $\pi^o$ 
decay background events and $24 \pm 2$ accidental coincidence background events is present in 
Fig.~\ref{fig:signal}. The residual $\pi^o$ decay background was estimated using the 
number of $\pi^o$ events with $\cos\theta_{12}  < -0.45$ and the residual accidental 
coincidence background was estimated using the number of $\pi$-$\pi$/$\mu$-$\pi$ events 
with $E_{\mathrm{sum}} > 150$~MeV.}
Their photon opening angle and individual energy spectra 
are shown in Fig.~\ref{fig:signal}.
The opening-angle spectrum is a broad continuum
and shows the opening angle cut of $\cos\theta_{12}   > -0.2$
we employed to remove the events from $\pi^o$ decay.
The individual energy spectrum is a broad continuum
with a low--energy cut-off at about 25~MeV 
due to the acceptance of the spectrometer.

The 370 two-photon events in Fig.~\ref{fig:signal}
are entirely consistent with
non-resonant capture in pionic deuterium.
Specifically,
using the measured branching ratio for non-resonant
double radiative capture on hydrogen of $(3.05 \pm 0.27 \pm 0.31) \times 10^{-5}$ \cite{Tr02},
and naively assuming the ratio of single radiative capture
to double radiative capture to be identical on a hydrogen target
and a deuterium target,
we would expect about $580 \pm 110$ events
from non-resonant $\pi^- d \rightarrow \gamma \gamma nn$.
In addition the measured angle and energy distributions
are in good agreement with our Monte Carlo simulations
of the non-resonant process using the theoretical model of
Beder \cite{Be79}.
The expected signature of dibaryon events,
a monoenergetic peak 
from the production process $\pi^ - d \rightarrow d^* \gamma $ 
and a three-body continuum
from the decay process $d^* \rightarrow n n \gamma$,
is not seen.

\section{Dibaryon Sensitivity}
To determine the detection efficiency for dibaryon events
we used a Monte Carlo computer program.
The program incorporated 
the detailed geometry of the RMC detector
and detailed interactions of the various particles.
Our program was based on the CERN GEANT3 package \cite{GEANT}
and is described in more detail in Ref.\ \cite{Wr98}.

The simulation was tested by measurements 
of the detector response with a liquid H$_2$ target.
Negative pion stops in hydrogen provide a well known source
of photon pairs from 
(i) $\pi^- p \rightarrow \pi^o n$ charge exchange 
followed by $\pi^o \rightarrow \gamma \gamma$ decay
and (ii) accidental $\gamma$-$\gamma$ coincidences
from multiple $\pi$ stops.
We found the energy-angle distributions from experiment and simulation
to be in good agreement. 
The absolute detection efficiencies
from experiment and simulation differed by 4\%
(the run-to-run variation of the spectrometer acceptance was $< \pm 6$\%).
This difference was attributed
to detector inefficiencies 
that were present in the measurement
but were absent in the simulation.
A multiplicative correction factor
$F = 0.96 \pm 0.06$ was therefore incorporated to account for this.

As described earlier, the $d^*$ signature 
is a monoenergetic $\gamma$-ray peak 
from dibaryon production, $\pi^- d \rightarrow d^* \gamma$, 
and a three-body continuum
from dibaryon decay,
$d^* \rightarrow n n \gamma$.
The production $\gamma$-ray energy
is approximately $E \sim M_{\pi^- d} - M_{d^*}$
and increases from $E \simeq 35$~MeV for $M_{d^*} = 1980$~MeV
to $E \simeq 90$~MeV for $M_{d^*} = 1920$~MeV.
The decay $\gamma$-ray spectrum
is peaked near the end-point energy, $E \sim M_{d^*} - 2 m_n$,
as typically the two neutrons carry 
only a small fraction of the available energy.
However the detailed shape of the three-body spectrum
is dependent on the $nn$-final state interaction
and consequently on the $d^*$ quantum numbers.

\begin{figure}
\begin{center} 
\includegraphics[width=0.7\hsize]{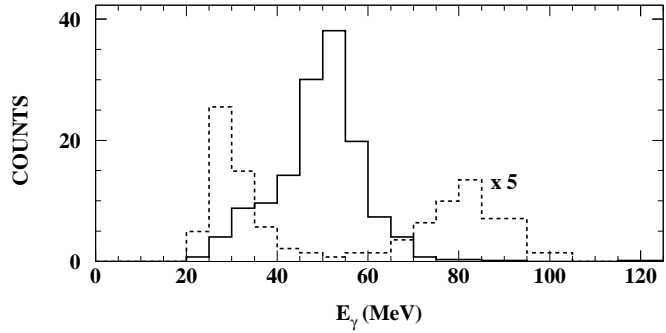}
\end{center}
\caption{Examples of the individual $\gamma$-ray energy spectra
for dibaryons with $M = 1956$~MeV (solid line) 
and $M = 1920$~MeV (dashed line) normalized using Eqn.~\ref{e:br}
and the corresponding branching ratio upper limit taken from Fig.~\ref{fig:brlimit}.
We assumed a three-body phase space distribution
for the $d^* \rightarrow \gamma n n$ decay $\gamma$-ray spectrum.}
\label{fig:simulation}
\end{figure}

\begin{figure}
\begin{center} 
\includegraphics[width=0.7\hsize]{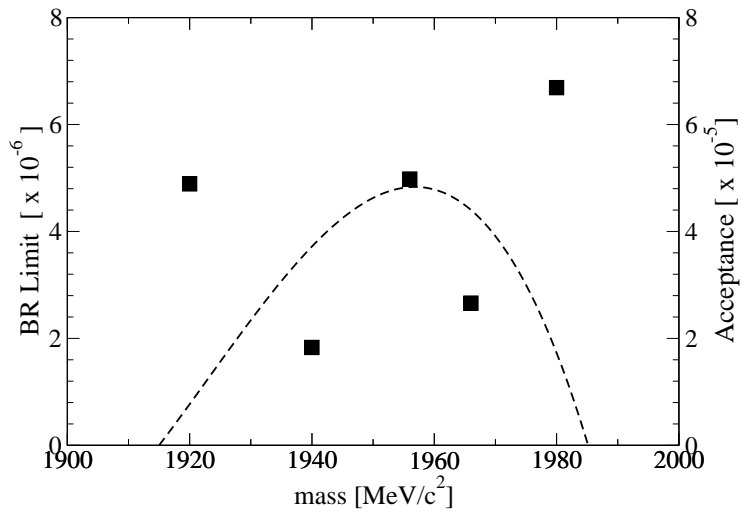}
\end{center}
\caption{The  90\%~C.L.~dibaryon branching ratio upper limit versus 
the $d^*$ mass (full squares and left-hand scale) and the Monte Carlo 
acceptance versus the $d^*$ mass (dashed line and right-hand scale).}
\label{fig:brlimit}
\end{figure}

Representative simulations of dibaryon signatures 
for $M_{d^*}$ = 1920 and 1956 MeV,
assuming a 3-body phase space distribution
for $d^* \rightarrow n n  \gamma$ decay,
are shown in Fig.~\ref{fig:simulation}.
Note that for $M_{d^*}$ = 1920 MeV
the production and decay $\gamma$-ray spectra are separated
whereas for $M_{d^*}$ = 1956 MeV
the production and decay $\gamma$-ray spectra overlap. 
Because the exact line shape of the decay $\gamma$-ray is not known,
we obtained our limits 
on $d^*$ production
by determining limits 
on the production $\gamma$-ray yield.
Specifically we fit the sum of 
a  polynomial function,
parameterizing the non-resonant background,
and a Gaussian peak,
accounting for $d^*$ production,
to the $\gamma$-$\gamma$ coincidence 
energy spectrum in Fig.~\ref{fig:signal}.
The Gaussian peak centroid was stepped
from $E = 35$ to $90$~MeV 
and the Gaussian peak width was fixed at the instrumental
resolution of the RMC detector
({\it i.e.} $\sigma = $4-5~MeV for $E = $40-80~MeV).
This procedure yielded a limit on the number of dibaryon events
as a function of dibaryon mass.
We then converted the event limit on dibaryon production
to a branching ratio limit on dibaryon production
via
\begin{equation}
\label{e:br}
B.R. = \frac{N_{d^*}} { N_{\pi^-} \cdot \epsilon\Delta\Omega \cdot F \cdot C} 
\end{equation}
where $N_{d^*}$ is the limit on the 
dibaryon events,
$N_{\pi^-}$ is the number 
of live-time corrected pion stops,
and $\epsilon\Delta\Omega \cdot F$ is the detector acceptance.
The appropriate acceptance was obtained 
using the Monte Carlo
for dibaryon production and assuming
a three-body phase space distribution
for $d^* \rightarrow n n \gamma$ decay (cf.~Fig.~\ref{fig:brlimit}).
As discussed earlier the factor $F$ accounts for detector inefficiencies
which are present in the experiment 
but are absent in the simulation. The factor $C = 0.85 \pm 0.01$ accounts
for the fraction of incident pions that stopped in deuterium
(see Wright {\it et al.}~\cite{Wr98} for details).

The resulting branching ratio upper limit on $d^*$ production
in $\pi^-$d capture was found to be
$BR < 6.7 \times 10^{-6}$ (90\% C.L.)
for $d^*$'s in the mass range of $1920$ to $1980$~MeV
and width of $< 10$~MeV
(see Fig.~\ref{fig:brlimit}).
In particular, we observed no evidence for
a narrow dibaryon of mass $M = 1956$~MeV
as claimed by Khrykin {\it et al.} \cite{Kh01}. 
However above and below this mass range,
our experimental sensitivity rapidly deteriorates
due to the energy cut-off in the spectrometer acceptance. 

\section{Summary}
In summary we have found no evidence 
for narrow dibaryon production in $\pi^-d$ capture.
Our upper limit on dibaryon production, $BR < 6.7 \times 10^{-6}$ (90\% C.L.),
is several orders of magnitude below 
the yield estimate of Gerasimov \cite{Ge98}.
Our null result is consistent 
with  the null result of Cal\'{e}n {\it et al.}\ \cite{Ca98}  
obtained from $pp$ bremsstrahlung measurements
using the WASA detector 
at the CELSIUS storage ring.



We wish to thank the staff 
of the TRIUMF laboratory
for their support of this work.
In particular we acknowledge the help
of Dr. Ren\'{e}e Poutissou on the data acquisition
and Dr. Dennis Healey on the cryogenic targets. 
In addition we thank Prof.~Sergo B.~Gerasimov 
for helpful and stimulating discussions,
Prof. David S.~Armstrong for carefully reading the manuscript
and the National Science Foundation (United States) 
and the Natural Sciences and Engineering Research Council (Canada) 
for financial support.


\newpage


\end{document}